\documentclass[12pt,tightenlines,eqsecnum,floats,aps,amsmath,amssymb,nofootinbib,prd,showpacs]{revtex4}

\usepackage{setspace}
\usepackage{amsmath,amssymb,amsfonts,amsthm}
\usepackage{graphicx}
\usepackage{enumerate} 
\usepackage{colordvi} 

\def\be{\begin{equation}}
\def\ee{\end{equation}}
\def\ba{\begin{eqnarray}}
\def\ea{\end{eqnarray}}
\def\bi{\begin{itemize}}
\def\ei{\end{itemize}}
\def\bnum{\begin{enumerate}}
\def\enum{\end{enumerate}}

\def\V{\mathcal{V}}

\def\p{\phi}
\def\H{{\cal H}}
\def\Hkg{\H_{\rm kin}^{\rm grav}}
\def\Hk{\H_{\rm kin}}

\def\Hkm{\H_{\rm kin}^{\rm matt}}

\def\f{\frac}

\def\lp{{\ell}_{\rm Pl}}
\def\l{\lambda}
\def\lo{\ell_o}
\def\bra{\langle}
\def\ket{\rangle}
\def\la{\lambda}
\def\sint{\textstyle{\int}}

\def\R{\mathbb{R}}

\def\Z{\mathbb{Z}}

\def\N{\mathbb{N}}
\def\t{\tau}
\def\T{\phi}

\def\frw{\text{FRW}}

\def\dd{\textrm{d}}

\def\tho{\Theta_0}
\def\thf{\Theta_1}

\def\cov{{\small  covariant} }
\def\can{{\small  canonical}}

\def\covtocan{{\small covariant $\to$ canonical} }
\def\cantocov{{\small canonical $\to$ covariant} }

\def\utw#1{\rlap{\lower1ex\hbox{$\sim$}}#1{}} 

\usepackage{enumerate}
\DeclareMathOperator*{\sgn}{sgn}

\usepackage{colordvi}
\newcounter{mnotecount}[section]

\newcommand{\comment}[1]{}
\begin{document}
\title{Vertex Expansion for the Bianchi I model}

\author{Miguel Campiglia\footnote{miguel@gravity.psu.edu}}

\author{Adam Henderson\footnote{henderson@gravity.psu.edu}}

\author{William Nelson\footnote{nelson@gravity.psu.edu} }
\affiliation{Institute of Gravitation and the Cosmos, Penn State University, State College, PA 16801, U.S.A. }

\begin{abstract}

A perturbative expansion of Loop Quantum Cosmological transitions amplitudes of Bianchi~I models is
performed. Following the procedure outlined in~\cite{ach1,ach2} for isotropic models,
it is shown that the resulting expansion can be written in the form of a series of amplitudes each with
a fixed number of transitions mimicking a spin foam expansion. This analogy is more complete than in the
isotropic case, since there are now the additional anisotropic degrees of freedom which play the role of
`colouring' of the spin foams. Furthermore, the isotropic expansion is recovered by integrating out
the anisotropies. 
\end{abstract}
\maketitle

\section{Introduction} \label{s1}

An important open problem in Loop Quantum Gravity (LQG)~\cite{alrev,crbook,crlrr,ttbook} is to obtain a well 
defined method of perturbatively computing its dynamics. The covariant approach given by Spin Foam Models
(SFM)~\cite{perezrev,jb-BF,crbook} provides an avenue to obtain such a method, however there are still several open issues regarding its precise
 relation to the Hamiltonian theory.  

If SFM and canonical LQG  are to be the covariant and canonical descriptions of a single quantum theory of gravity, one
should be able to derive one from the other. We are still far from rigorously establishing such connection,
however important progress has arisen in recent years. These include, in the \covtocan direction,
the derivation of the LQG Hilbert space as well as the spectra of geometrical
operators~\cite{newlook,engleper,dingrov}, and, in the \cantocov direction, the extension of the EPRL
amplitude to arbitrary valent spin foams vertices thereby allowing general histories of graphs~\cite{kkl}. 
The latter direction leads also to the picture of regarding spin foams as spin networks histories~\cite{rr}. 
This interpretation however, has not gone beyond the heuristic level (as far the full four dimensional theory 
is concerned; in the three dimensional case the connection between the canonical and covariant descriptions is well established~\cite{np}). 

Recently, the \cantocov direction was analyzed at the symmetry reduced level of homogeneous and isotropic
cosmology~\cite{ach1,ach2}. Here we extend that analysis and consider a non-isotropic cosmological model. The
additional degrees of freedom allow for a richer discussion than in the isotropic case. In particular, besides
the `vertex expansion' present in the Friedman-Robertson-Walker (FRW) case, there are additional sums over the
extra parameters, which are interpreted as giving a `colouring' of the graph, thus strengthening the analogy with spin foams.

The aim here is to obtain a `sum over histories' description within 
Loop Quantum Cosmology (LQC)~\cite{mblrr}. Our construction is then different from that of `spinfoam cosmology' \cite{rv1,brv},
 where cosmological amplitudes are obtained using SFM as the starting point. An interesting question which we do not address here
 is whether there is any precise relation among the two constructions. Let us also mention that the idea of looking for a \mbox{\can-\cov}
 connection at the homogeneous level has appeared before in the context of Plebanksi theory~\cite{npv}; such approach could shed light
 into the previous question of relating  `cosmological spinfoams' with `spinfoam cosmologies'.

The paper is organized as follows. In Section~\ref{s2} we introduce the model we will work with, namely the quantum Bianchi~I
cosmology with a massless scalar field, as obtained by Ashtekar and Wilson-Ewing in~\cite{awe}. In Section~\ref{s3} we construct
the sum over histories description of the model. We outline  how individual amplitudes are to be calculated and illustrate the
procedure for a simple history. In Section \ref{s4} the vertex expansion of the FRW~\cite{ach1} model is recovered by integrating out
 the anisotropies of our vertex expansion. We finish the paper with a discussion in Section~\ref{s5}.

\section{Loop Quantum Cosmology of the Bianchi~I model} \label{s2}

We are interested in the Bianchi~I cosmological model, which is the 
simplest non-isotropic homogeneous cosmology, coupled with a massless scalar field $\phi$.
As in the isotropic case, one can fix a fiducial 3-metric $ds_0^2$ and choose Cartesian coordinates on the spacial slice
 such that ${\rm d}s_0^2= {\rm d}z_1^2 + 
{\rm d}z_2^2 + {\rm d}z_3^2$. The physical 3-metric is then determined by three  directional scales factors, $a_1,a_2$ and $a_3$,
\be
{\rm d}s^2 =  a_1^2 ~ {\rm d}z_1^2 + a_2^2 ~ {\rm d}z_2^2 + a_3^2 ~ {\rm d}z_3^2~.
\ee
The Hamiltonian analysis requires one to choose a fiducial cell $\V$, for which we take the rectangular prism  $0 \leq z_i
\leq L_i$, for each direction $i=1,2,3$. The physical volume of the cell is then given by $V= |a_1 a_2 a_3| L_1 L_2 L_3$.
Note that the choice of the fiducial cell $\V$ (i.e.\ the choice of $L_1$, $L_2$ and $L_3$)
 is arbitrary, and one has to ensure that physical results are insensitive to that choice (see \cite{awe} for further discussion).

When one goes to the quantum theory \cite{awe}, it is convenient to work with a new set of variables, $(\la_1,\la_2,v)$,
defined by 
\ba \label{lambda}
\la_1 & := & \frac{\sgn(a_1) \sqrt{|a_2 a_3| L_2 L_3}}{(4 \pi \gamma \lp^2 \lo )^{1/3}}~, \\
\la_2 & := & \frac{\sgn(a_2) \sqrt{|a_3 a_1| L_3 L_1}}{(4 \pi \gamma \lp^2 \lo )^{1/3}}~, \\
v & := & \frac{\sgn(a_1 a_2 a_3) |a_1 a_2 a_3| L_1 L_2 L_3}{2 \pi \gamma \lp^2 \lo } = 2 \l_1 \l_2 \l_3.
\ea
Here $\lo$ is the square root of the `area gap' $\Delta =4\sqrt{3}\pi
\gamma\, \lp^2$, $\gamma$ is the Barbero-Immirizi parameter, and $\l_3$ is defined in a similar way as $\l_1$ and $\l_2$. In
this representation, the gravitational Hilbert space $\Hkg$ consists of functions
$\Psi(\la_1,\la_2,v)$ with support on a
countable number of points and with finite norm $||\Psi||^2 :=
\sum_{\la_1,\la_2,v}\,|\Psi(\la_1,\la_2,v)|^2 <\infty$. The matter Hilbert space is the standard
one: $\Hkm = L^2(\R, \dd\p)$. The total kinematical Hilbert space is thus a tensor product $\Hk=\Hkg \otimes
\Hkm$ and, as usual in LQC, the dynamics of the system are encoded in the constraint equation
\be  -\,C \Psi \equiv \partial^2_\p \Psi + \Theta \Psi = 0~,
\ee
where $\Theta$ is a symmetric operator that acts on the gravitational part. As in \cite{awe}, we restrict attention to the
`positive octant' ($v,\l_1,\l_2 \geq 0$). The action of $\Theta$ takes the form,  
\begin{align} \label{qham6} \left(\Theta \Psi \right)(\l_1,\l_2,v) =& \f{\pi G}
{4}\sqrt{v}\Big[(v+2)\sqrt{v+4}\,\Psi^+_4(\l_1,\l_2,v) - (v+2)\sqrt
v\, \Psi^+_0( \l_1,\l_2,v)\nonumber \\& -(v-2)\sqrt v\,
\Psi^-_0(\l_1,\l_2,v) + (v-2)
\sqrt{|v-4|}\,\Psi^-_4(\l_1,\l_2,v)\Big], \end{align}
where $\Psi^\pm_{0,4}$ are defined as:
\begin{align} \label{qham7}\Psi^\pm_4(\l_1,\l_2,v)=& \:\Psi
\left(\f{v\pm4}{v\pm2}\cdot\l_1,\f{v\pm2}{v}\cdot\l_2,
v\pm4\right)+\Psi\left(\f{v\pm4}{v\pm2}\cdot\l_1,\l_2,
v\pm4\right)\nonumber\\& +\Psi\left(\f{v\pm2}{v}\cdot\l_1,
\f{v\pm4}{v\pm2}\cdot\l_2,v\pm4\right)+\Psi
\left(\f{v\pm2}{v}\cdot\l_1, \l_2,v\pm4\right)\nonumber
\\&+\Psi\left(\l_1,\f{v\pm2}{v}\cdot\l_2,
v\pm4\right)+\Psi\left(\l_1,\f{v\pm4}{v\pm2}\cdot\l_2,v\pm4\right),
\end{align}
and
\begin{align} \label{qham8} \Psi^\pm_0(\l_1,\l_2,v)=
& \:\Psi\left(\f{v\pm2}{v}\cdot\l_1, \f{v}{v\pm2}\cdot\l_2,v\right)
+\Psi\left(\f{v\pm2}{v}\cdot\l_1,\l_2,v \right)\nonumber \\&
+\Psi\left(\f{v}{v\pm2}\cdot\l_1,\f{v\pm2}{v}\cdot\l_2,v\right)+\Psi\left(\f{v}{v\pm2}\cdot\l_1,\l_2,v\right)\nonumber
\\& +
\Psi\left(\l_1,\f{v}{v\pm2}\cdot\l_2,v\right)+\Psi\left(\l_1,\f{v\pm2}{v}
\cdot\l_2,v\right)\, .\end{align}
As noted in \cite{awe}, the operator preserves the subspaces $\H_\epsilon \subset \Hkg$ of states whose support lies on the lattice
\be
v = \epsilon + 4 \Z~.
\ee
These superselection sectors have the same form as in the isotropic case \cite{acs}, and, as done there, we will restrict
to the physically interesting $\epsilon = 0$ sector (the one that contains the classically singular value  $v=0$).  The space we
finally work with is  the space of vectors with support on the positive octant and the $\epsilon=0$ lattice, which we denote
by $\H^{+++}_{0}$.

We now introduce a different representation of the space $\H^{+++}_{0} \subset \Hkg$, by changing coordinates
$(\la_1,\la_2,v) \to (n,{\bf x})$ where
\ba \label{eq:intertwiner}
n  & := & \frac{1}{4} v \in \N~, \\
{\bf x}   =  (x_1,x_2) & := & (\log \la_1,\log \la_2) \in \R^2~.
\ea
In this representation, states are described by functions $\Psi(n,{\bf x})$, which again have support on a countable
number of points and have a finite norm $||\Psi||^2 = \sum_{{\bf x},v}\,|\Psi({\bf x},v)|^2<\infty$.
They represent the components of the state in a
$\{ |n,{\bf x} \ket \}$ basis, which is characterized by eigenvalues of  $\widehat{\la}_1,\widehat{\la}_2$ and
$\widehat{V}=4 \pi \gamma \lp^2 \lo \widehat{\la}_1\widehat{\la}_2\widehat{\la}_3$ as follows 
\ba
\widehat{V} |n,{\bf x} \ket & = & 8 \pi \gamma \lp^2 \lo n |n,{\bf x} \ket~, \\
\widehat{\la}_i |n, {\bf x} \ket & = & \log x_i |n,{\bf x} \ket , \quad i=1,2~,
\ea
with their normalization given by the product of Kronecker deltas:
\be
\bra n', {\bf x}'| n,  {\bf x} \ket = \delta_{n,n'} \delta_{{\bf x},{\bf x}'}.
\ee

In the subsequent sections, we will regard $\H^{+++}_{0}$ as a tensor product of the `volume' factor and the `anisotropy' factor:
\ba
\H^{+++}_{0} & = &\H_V \otimes \H_\la~, \\
|n, {\bf x} \ket & = & |n \ket \otimes | {\bf x} \ket~.
\ea
Within this splitting, $\Theta$ is expressed as a sum of the tensor product of operators acting on $\H_V$ and $\H_\la$,
\be \label{theta}
\Theta = \sum_n | n+1 \ket \bra n| \otimes \Theta_{\left(n+1\right) n} + | n \ket \bra n| \otimes \Theta_{nn} +
 | n-1 \ket \bra n| \otimes \Theta_{\left(n-1\right) n}~.
\ee
The form of the operators acting on the anisotropy factor is quite simple: they are composed of translations in the
 ${\bf x}$ plane of lengths
\be \label{stepsize}
a^\pm_n  :=\log \frac{2n\pm 1}{2n}~,
\ee
which depend on the volume $n$. If we write the operator generating the translation by $a_n$ in the $x_1$ direction as
\be
\left(e^{i a_n p_1} \Psi \right) (n, x_1 ,x_2):= \Psi(n, x_1 +a_n ,x_2)~,
\ee
and similarly for translations in the $x_2$ direction, the operators acting on $\H_\la$ take the form
\be \label{diag}
 \Theta_{n n}   =   2 \pi G \left( n (2n + 1)\left[ \cos a^+_n p_1 + \cos a^+_n p_2 + \cos (a^+_n p_2-a^+_n p_1)
\right]+n(2n-1) [ a^+_n \to a^-_n ] \right)~,
\ee 
\be \label{offdiag}
\Theta_{n\pm1\; n}  =   - \pi G \sqrt{n(n \pm 1)}(2n \pm 1)  \left( e^{-i a^\pm_n p_1} + e^{i a^\pm_{n\pm 1/2} p_1}
+ e^{-i a^\pm_n p_1} e^{i a^\pm_{n\pm 1/2} p_2} + p_1 \leftrightarrow p_2 \right)~.
\ee
One can easily verify that the action of Eq.~(\ref{theta}), when written in terms of the original representation, reproduces 
Eq.~(\ref{qham6}).

Let us conclude this section by mentioning a remarkable property of the $\Theta$ operator. As found in \cite{awe}, one can recover the
FRW cosmology by `integrating out' the anisotropies of the Bianchi I model. Specifically, it was shown that there is a
projection map from the Bianchi~I space to the FRW space defined by
\be
\sum_{{\bf x}} \Psi(n,{\bf x};\T) = \Psi^\frw (n ; \T)~,
\ee
in which the $\Theta$ operator is mapped to the $\Theta^\frw$ operator of the FRW model\footnote{See~\cite{Nelson:2009yn}
for an alternative projection that produces isotropic states, but not the $\Theta^\frw$ associated
with the $\nu$-quantization procedure.} namely, 
\be
\sum_{{\bf x}} \Theta \Psi(n,{\bf x} ; \T) = \Theta^\frw \Psi^\frw (n ; \T).
\ee
We will later see how this projections holds order by order in the vertex expansion.

\section{Sum over histories} \label{s3}

As in \cite{ach1,ach2}, the natural object on which to construct a sum over histories expansion is the physical inner product between
 `initial' $| [n_i, {\bf x}_i, \p_i] \ket$ and `final' $| [n_f,{\bf x}_f, \p_f] \ket$ physical states. This inner product is constructed 
from a group averaging formula involving the kinematical states $|n_i, {\bf x}_i, \p_i \ket$ and $|n_f, {\bf x}_f, \p_f \ket$,
\be \label{physip-0}
 ([n_f,{\bf x}_f, \p_f] , [n_i, {\bf x}_i, \p_i] ) =
 2 \bra n_f, {\bf x}_f, \p_f | \sint_{-\infty}^{\infty}
 \dd{\alpha} \,e^{i \alpha C} \ |{p}_{\phi}|\, | n_i, {\bf x}_i,\p_i \ket,
\ee
(the $|{p}_{\phi}|$ term is there so that the normalization agrees with the one used in \cite{ach2}).
As in the FRW case \cite{ach1,ach2}, a  key
simplification comes from the fact that the constraint $C$ is a
sum of two commuting pieces that act separately on $\Hkm$ and
$\Hkg$. Consequently, the integrand of Eq.~(\ref{physip-0}) splits into a matter and gravitational factors:
\be \label{integrandsplit}
2 \bra n_f, {\bf x}_f, \p_f |  \,e^{i \alpha C} \ |{p}_{\phi}|\, | n_i, {\bf x}_i,\p_i \ket  = 2\,\bra \p_f |
e^{i \alpha p_{\p}^2 }\, |p_{\p}|  | \p_i \ket \bra n_f, {\bf x}_f | e^{-i \alpha \Theta} | n_i, {\bf x}_i \ket.
\ee
The matter part can be easily evaluated as,
\be
\label{aphi} 2\,\bra \p_f |  e^{i \alpha
p_{\p}^2 }\, |p_{\p}|  | \p_i \ket\,=\, 2\,\sint \frac{\dd p_\p}{2 \pi}
e^{i \alpha p_\p^2}\, e^{i p_\p (\p_f-\p_i)}\, |p_\p|~.
\ee
The non-triviality of Eq.~(\ref{physip-0}) lies in the gravitational part, $\bra n_f, {\bf x}_f | e^{-i \alpha \Theta} | n_i, {\bf x}_i \ket$. 
 Following the strategy depicted in \cite{ach1}, we will express such term as a sum over histories. This can be achieved by observing that
 the term has the form of a matrix element of a fictitious evolution operator $e^{-i \alpha \Theta}$, with $\Theta$ playing the role of
 Hamiltonian and $\alpha$ that of time. 

Once the gravitational factor is written as a sum over histories, the idea is to perform the integral over $\alpha$ for each history
 separately, obtaining at the end a sum over histories expansion of the physical inner product. These steps will be discusses in the
 following subsections.

\subsection{Sum over histories for the gravitational amplitude}

To construct a `sum over histories' expansion of the gravitational amplitude $\bra n_f, {\bf x}_f | e^{-i \alpha \Theta} |
n_i, {\bf x}_i \ket$, one would proceed with a Feynman-like procedure of dividing the `time' $\alpha$ into $N$ steps of length
$\epsilon = \alpha/N$, inserting a complete basis in between each factor, and finally taking the $N \to \infty$ limit. In~\cite{ach2}
it was shown (in the FRW context, but the result is generic for any discrete labeled basis) that the resulting limit is equivalent to a specific perturbative expansion of the `evolution' operator under study. We will use this result here to construct the sum over histories directly from the perturbation series.

The starting point in such a derivation is to write the fictitious Hamiltonian $\Theta$ as an `unperturbed part'
$\tho $ plus a `perturbation' $\thf$,
\be\label{eq:diag_offdiag}
\Theta=\tho + \thf~.
\ee

In a spin network/spin foam picture, the above splitting would correspond to a graph preserving piece, $\tho$, plus the
 remaining graph changing part, $\thf$. In our case, we choose to interpret the label $n$ as containing the information 
of the `graph' , and the remaining ${\bf x}$ label as the colouring of the graph. Thus, $\tho$ and $\thf$ are respectively
 diagonal and off-diagonal in $n$. In the tensorial notation used in Eq. (\ref{theta}), these operators are given by,
\ba 
\tho & = & \sum_n   | n \ket \bra n| \otimes \Theta_{n n}~,  \\
\thf & = & \sum_n | n+1 \ket \bra n| \otimes \Theta_{\left(n+1\right) n} + | n-1 \ket \bra n| \otimes \Theta_{\left(n-1\right) n}~.
\ea

The construction now follows as in the FRW case \cite{ach2}, where the same label was used to trigger the transitions.
Using standard perturbation theory in the interaction picture,  the transition amplitude is be written as
\begin{align} \label{exp1-1}
\bra n_f,{\bf x}_f | \ e^{-i \alpha \Theta}\ | n_i, {\bf x}_i \ket & =  \bra n_f,{\bf x}_f| \bigg[ \sum_{M=0}^{\infty} (-i)^M
 \int_{0}^{\alpha} \dd \t_{M}
\ldots \int_{0}^{\t_2}\! \dd \t_1  \nonumber \\
   & e^{-i (\alpha-\t_M)\tho} \thf e^{-i (\t_M-\t_{M-1})\tho} \thf \ldots 
e^{-i (\t_2-\t_{1})\tho} \thf  e^{-i \t_{1}\tho} \bigg]  | n_i, {\bf x}_i\ket~.
\end{align}
The $M$-th term of the sum generates all histories with $M$ transitions. These histories are obtained by inserting $M-1$
identities in the form $\mathbf{1}=\sum_{n_m} \left(| n_m \ket \bra n_m | \right) \otimes \mathbf{1}_{\la}$, $m=1,\ldots,
M-1$ next to each $\Theta_1$ factor. This results in a sum over a sequence of volumes $(n_0,n_1,\ldots,n_M)$ (with $n_0 \equiv n_i$
and $n_M \equiv n_f$ held fixed), given by
\be \label{sohg}
\bra n_f,{\bf x}_f| \ e^{-i \alpha \Theta}\ | n_i, {\bf x}_i\ket = \sum_{M=0}^\infty \Big[\sum_{n_{M-1},\ldots,n_{1}}
\bra {\bf x}_f | A(n_M,\ldots,n_0;\alpha) |{\bf x}_i \ket \Big]~,
\ee
where
\begin{align} \label{amphg}
& A(n_M,\ldots,n_0;\alpha)  :=  (-i)^M \int_{0}^{\alpha} \dd \t_{M} \ldots  \int_{0}^{\t_2}\! \dd \t_1 \;  A(n_M,\ldots,n_0;
\t_M,\ldots,\t_1;\alpha)~, \\ 
& A(n_M,\ldots,n_0;\t_M,\ldots,\t_1;\alpha)  :=  A_{n_M}(\alpha-\t_M) V_{n_M n_{M-1}}A_{n_{M-1}}(\t_M-\t_{M-1}) \ldots
V_{n_1 n_0}A_{n_0}(\t_1)~,
\label{amphgt} 
\end{align} 
with  $A_n\left(\tau\right)$ and $V_{n'n}$  defined as,
\ba 
&A_n(\t) &:= e^{-i \t \Theta_{nn}}  \label{A-1} \\ 
&V_{n' n}& : =  \left\{ \begin{array}{ll} \Theta_{n' n} \label{V-1}
& \quad n' \neq n \\
0 & \quad n' = n.
\end{array}\right.
\ea

Note that all the factors in Eqs.~(\ref{amphg} - \ref{V-1}) are operators on $\H_\la$ , whilst the actual amplitude,
Eq.~(\ref{sohg}), involves matrix elements of the operator defined by Eq.~(\ref{amphgt}). Note also that the only
sequences entering in the sum are such that $n_{m}=n_{m-1} \pm 1, m=1,\ldots,M$. In particular, for $M$ fixed, there
is a finite number of terms.

The construction above has the same form as in the FRW case. The distinction however lies on the fact that there are
additional degrees of freedom, given by the anisotropies ${\bf x}$. However, in the description given so far, intermediate
anisotropies do not appear since they are implicitly `summed over'. To make these additional sums explicit, we insert
identities in the form $\mathbf{1}_\la= \sum_{{\bf x}_m}| {\bf x}_m \ket \bra {\bf x}_m |$  and 
$\mathbf{1}_\la= \sum_{{\bf y}_m}| {\bf y}_m \ket \bra {\bf y}_m |$ to the right and left of the $A_{n_m}(\t_{m+1}-\t_m)$ operators in
Eq.~(\ref{amphgt}). The gravitational amplitude (\ref{exp1-1}) then takes the form,

\begin{align}\label{diagram}
  & \bra n_f,{\bf x}_f| \ e^{-i \alpha \Theta}\ | n_i, {\bf x}_i\ket = \nonumber \\ 
& \sum_{M=0}^\infty  \sum_{n_{M-1},\ldots,n_{1}} \sum_{\substack{{\bf x}_1,\ldots,{\bf x}_M\\{\bf y}_0,\ldots,{\bf y}_{M-1}}}
 A({\bf y}_M, n_M, {\bf x}_M, {\bf y}_{M-1}, n_{M-1}, \ldots, {\bf y}_0, n_0, {\bf x}_0 ; \alpha)
\end{align}
where now we have, on top of the `graph history' sum (given by the volume sequence), a sum over all possible `colourings' for
 each `graph history'. The amplitude for such history is given by
\ba \label{amphg2}
&&A({\bf y}_M, n_M, {\bf x}_M, {\bf y}_{M-1}, n_{M-1}, \ldots, {\bf y}_0, n_0, {\bf x}_0 ; \alpha) :=\\
&&\nonumber  (-i)^M \int_{0}^{\alpha} \dd \t_{M} \ldots  \int_{0}^{\t_2}\! \dd \t_1 \;  A({\bf y}_M, n_M, {\bf x}_M,
 {\bf y}_{M-1}, n_{M-1}, \ldots, {\bf y}_0, n_0, {\bf x}_0  ;
\t_M,\ldots,\t_1;\alpha)~,
\ea
where
\ba \label{ampanis}
&&A({\bf y}_M, n_M, {\bf x}_M, {\bf y}_{M-1}, n_{M-1}, \ldots, {\bf y}_0, n_0, {\bf x}_0  ;
\t_M,\ldots,\t_1;\alpha) :=\\
&&\nonumber  A_{n_M {\bf y}_M {\bf x}_M }(\alpha-\t_M)V_{n_M {\bf x_M} n_{M-1} {\bf y}_{M-1}} A_{n_{M-1} {\bf y}_{M-1}
 {\bf x}_{M-1}}(\t_M-\t_{M-1}) \ldots
V_{n_1  {\bf x_1} n_{0} {\bf y}_{0}}A_{n_0}(\t_1)~,
\ea
and
\ba
&A_{n {\bf y} {\bf x}}(\tau) & := \bra {\bf y} | A_n(\tau) | {\bf x} \ket \label{A-2} \\
&V_{n' {\bf x} n {\bf y}} &:= \bra {\bf x}| V_{n'n} | {\bf y} \ket \quad \label{V-2}.
\ea
Note that, in spite of their appearance, the sums over intermediate anisotropies in (\ref{diagram}) are well defined.
Although in principle the anisotropies can take any real value, in practice only a countable subset of the real numbers is
involved in the sum (the amplitude vanishes elsewhere).  More details on this point are given in Section~\ref{histamp}.

\subsection{Vertex expansion of the physical inner product}

\begin{figure}
 \includegraphics{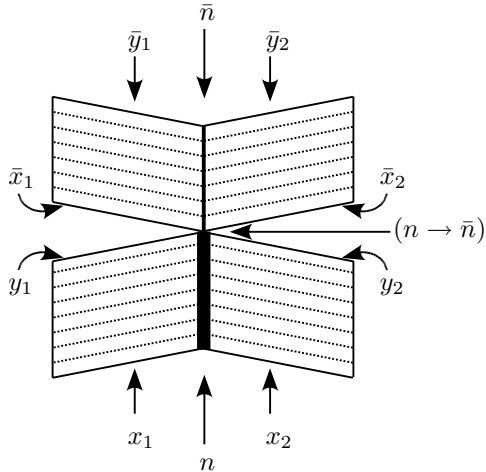}
 \caption{\label{fig:2} The history of the spin network state i.e.\ the spin foam analogue of our vertex expansion,
with $M=1$ is represented pictorially. The initial state has volume $n$ and anisotropies $x_i$. The state then `evolves', keeping $n$ constant but allowing the anisotropies to vary. Eventually there is a transition to a new volume, $\bar{n}$. The anisotropy before and after the transition are $y_i$ and $\bar{x}_i$ respectively. The final state has labels $\bar{n}$ and $\bar{y}_i$. 
 The dashed lines indicate that the `spin' labels ($x_i$) evolve along each of the constant volume pieces. The amplitude for such process, in the notation of Eq. (\ref{amppath}), is given by $A(\bar{{\bf y}}, \bar{n}, \bar{{\bf x}},  {\bf y}, n, {\bf x})$.}
\end{figure}

We now use the above construction to obtain an expansion for the physical inner product, Eq.~(\ref{physip-0}). First, one
rewrites the integrand of Eq.~(\ref{physip-0}) as in Eq.~(\ref{integrandsplit}) and then the gravitational factor
is written using the expansion given in Eq.~(\ref{diagram}).
One then interchanges the integral over $\alpha$ with the sums over $M$ and the intermediate labels, to arrive at a `sum over histories' expansion of the
physical inner product,
\ba 
 ([n_f,{\bf x}_f, \p_f] , [n_i, {\bf x}_i, \p_i] ) & = & 
 \sum_{M=0}^\infty  \sum_{n_{M-1},\ldots,n_{1}} \sum_{\substack{{\bf x}_1,\ldots,{\bf x}_M\\{\bf y}_0,\ldots,{\bf y}_{M-1}}}A({\bf y}_M, n_M, {\bf x}_M, {\bf y}_{M-1}, n_{M-1}, \ldots, {\bf y}_0, n_0, {\bf x}_0) \nonumber \\
&=& \sum_{M=0}^\infty  \sum_{n_{M-1},\ldots,n_{1}} A(n_M,\ldots,n_0;{\bf x}_i,{\bf x}_f;\p_i,\p_f) \label{physip-1}
\ea
where

\begin{align} \label{amppath}
 A({\bf y}_M, n_M, {\bf x}_M,  {\bf y}_{M-1}, n_{M-1}, & \ldots,  {\bf y}_0, n_0, {\bf x}_0) :=2\,\sint \frac{\dd p_\p}{2\pi} \;
 e^{i p_\p (\p_f-\p_i)}\, |p_\p| \sint_{-\infty}^{\infty}
 \dd{\alpha} \, \\
 & e^{i \alpha p_\p^2} A({\bf y}_M, n_M, {\bf x}_M, {\bf y}_{M-1}, n_{M-1}, \ldots, {\bf y}_0, n_0, {\bf x}_0 ;
\alpha)~.\nonumber 
\end{align}
\begin{figure}
 \includegraphics{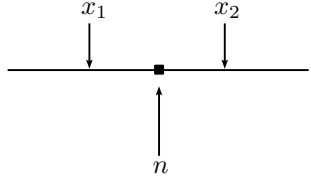}
 \caption{\label{fig:1} The Bianchi~I model can be written in terms of three parameters $(n, x_1, x_2)$. $n$ dictates the
volume, whilst $x_1$ and $x_2$ are analogous to the spin labels
of LQG and are represented here as two edges. Note that this is not intended to be a true LQG graph,
rather it is a pictorial description of the degrees of freedom that our states have.} 
\end{figure}
Pictorially, we can represent the expansion as follows. First, we represent the gravitational ket $| n, {\bf x } \ket$
as depicted in Fig.~\ref{fig:1}. A `history' with one transition in $n\rightarrow \bar{n}$ is then represented in Fig.~\ref{fig:2}.

Note that here the analogue
to spin foams is not exact, since in a spin foam, the spin labels on each face of the triangulation are constant. In the expansion
derived here there is non-trivial dynamics for the `spins', ${\bf x}_i$, even with a fixed `graph' $n$. In general there are
two distinct labels for a face, those at the beginning, ${\bf x}_i$ and those at the end, ${\bf y}_i$, and  ${\bf y}_i \neq
{\bf x}_i$. Also, in a 
spin foam there is no restriction on the spins across a vertex, whereas in our case ${\bf y}_i$ and ${\bf x}_{i+1}$
(the `spin' labels on either side of the $(i+1)^{\rm th}$ vertex ) are very closely related, by the form of
Eq.~(\ref{offdiag}). Thus, although the analogue is not complete, the form of the expansions are qualitatively the
same.

\subsection{Histories amplitudes} \label{histamp}

We now discuss how path amplitudes appearing in Eq.~(\ref{physip-1}) can be calculated and illustrate the procedure
in the simplest case.  The first step is to evaluate the matrix element $\bra {\bf x}_f |A(n_M,\ldots,n_0; \tau_M,
\ldots, \tau_1;\alpha)|{\bf x}_i \ket$ of the operator defined in Eq.~(\ref{amphgt}).  This operator consist of compositions
 of operators $A_n(\t)$ and $V_{n' n}$, given respectively in Eq.~(\ref{A-1}) and Eq.~(\ref{V-1}), which themselves are
constructed from translations in the  ${\bf x}$ plane.  Because of the translation invariance, we will consider the matrix
elements between  $\bra {\bf x}|$ and $|0 \ket$; the original matrix element is then recovered by the substitution
${\bf x} \to {\bf x}_f-{\bf x}_i$.

Let us discuss the structure of the operators in more detail.  As already noted, $V_{ n' n}$ vanishes unless $n'=n\pm 1$, in which
case it is given by Eq.~(\ref{offdiag}).  It consist of an overall factor times six simple shifts involving lengths of $a^\pm_n$ and
 $a^\pm_{n \pm 1/2}$.  The $A_n(\t)$ operator is the exponentiation of ($-i \t$ times) the operator given in
Eq.~(\ref{diag}).  It can be factored into two terms,
\be
A_n(\t)=A^+_n(\t)A^-_n(\t)~,
\ee
where
\be
A^\pm_n(\t)=e^{-i \t 2 \pi G \, n (2n \pm 1)\left[ \cos a^\pm_n p_1 + \cos a^\pm_n p_2 + \cos (a^\pm_n p_2-a^\pm_n p_1) \right]}~,
\ee
which only involves shifts with step-size $a^\pm_n$.  The total operator $A(n_M,\ldots,n_0;\t_M,\ldots,\t_1;\alpha)$ is then a product
 of operators involving shifts of lengths $a^\pm_j$ with $j=n_m$ or $j=n_m \pm 1/2$ ($m=0,\ldots,M$).
 As a result the matrix element between $\bra {\bf x} |$ and $|0 \ket $ vanishes unless
${\bf x}$ lies in the `lattice' generated by the $a^\pm_j$ steps. The computation simplifies by selecting among the
$a^\pm_j$'s, a set of independent (i.e. incommensurate) lengths that generate the `lattice'. 

Let us illustrate the situation by considering the simplest path, namely the $M=0$ case. In this case the amplitude is given by
 the matrix element $\bra {\bf x} |A_n(\t) |0 \ket$.   As already noticed, this operator contains two step-sizes, $a^+_n$ and
 $a^-_n$, and so the nontrivial matrix elements occur whenever ${\bf x}$
lies in the lattice~\footnote{Notice that the lattices we will referring to are not the usual ones (as for instance a square
 lattice $a \Z^2 \subset \R^2$) where points form a grid. Rather they fill in the entire plane. For instance one can show that
 the points of the lattice defined by   Eq.~(\ref{lattice}) form a dense subset of $\R^2$ (because $a^+_n$ and $a^-_n$ are
 incommensurate numbers).},
\be \label{lattice}
{\bf x} = {\bf k}^+ a^+_n+{\bf k}^- a^-_n \ ; \qquad   {\bf k}^+ = (k^+_1,k^+_2) \in \Z^2, \; {\bf k}^- = (k^-_1,k^-_2)\in \Z^2~.
\ee
From the definition of $a^+_n$ and $a^-_n$, Eq.~(\ref{stepsize}), one can check that these two numbers are always incommensurate,
and so they form an independent set of generators of the lattice.  That is to say, a point in the lattice defined by Eq.~(\ref{lattice}) is
uniquely decomposed into its $a^+_n$ and $a^-_n$ components, i.e., if  ${\bf k}^+ a^+_n+{\bf k}^- a^-_n = {\bf l}^+ a^+_n+
{\bf l}^- a^-_n$ then ${\bf k}^\pm={\bf l}^\pm$. 

Thus, the kets $|{\bf x}= {\bf k}^+ a^+_n+{\bf k}^- a^-_n \ket$, form a basis of the subspace of vectors which give a
non-vanishing matrix element.  This space has the structure of a tensor product of two copies of $L^2(\Z^2)$,
\be
|{\bf x}= {\bf k}^+ a^+_n+{\bf k}^- a^-_n \ket~= |{\bf k}^+ \ket \otimes |{\bf k}^- \ket 
\ee
where $|{\bf k}^+ \ket$ and $|{\bf k}^- \ket$ are viewed as basis of two abstract $L^2(\Z^2)$ spaces.

Viewed in this way, the $A^\pm_n(\t)$ operators act separately on each $L^2(\Z^2)$ factor:
\be \label{mea}
\bra {\bf x}= {\bf k}^+ a^+_n+{\bf k}^- a^-_n |A_n(\t) | 0 \ket = \bra {\bf k}^+|A^+_n(\t)|0 \ket \bra {\bf k}^-|A^-_n(\t)|0 \ket~.
\ee
Each factor is the matrix element of a (translation invariant) operator in a single $L^2(\Z^2)$ space, and so can
be evaluated by taking the Fourier transform.  Using Eq.~(\ref{diag}) one finds 
\be \label{apm}
\bra {\bf k}^\pm|A^\pm_n(\t)|0 \ket=\int_0^{2 \pi}\frac{\dd{\theta_1}}{2 \pi} \int_0^{2 \pi}\frac{\dd{\theta_2}}{2 \pi}
e^{i k^\pm_1 \theta_1+i k^\pm_2 \theta_2}e^{-i 2\pi G n(2n \pm 1) \t \left(\cos \theta_1 + \cos \theta_2 + \cos
(\theta_2-\theta_1) \right)}~.
\ee

In the general case of a path with $M$ transitions and $g$ independent generators, the space of vectors giving a non-vanishing matrix elements will have now the structure of a tensor
product of $g$ copies of $L^2(\Z^2)$. Because the vertex $V_{n' n}$ is the sum of six terms, one will generically have a
total of $\sim6^M$ terms, each of them involving $g$ Fourier integrals of the type described above.  The expressions for
these integrals can be directly read off from Eqs.~(\ref{diag}) and (\ref{offdiag}).

After the evaluation of the matrix element $\bra {\bf x}_f |A(n_M,\ldots,n_0; \tau_M,\ldots, \tau_1;\alpha)|{\bf x}_i \ket$,
 one has to perform the $\t$ integrals in Eq.~(\ref{amphgt}), and the $p_\p$ and $\alpha$ integrals in Eq.~(\ref{amppath}).
In our $M=0$ case, there are no $\t$ integrals to perform, and the
 $\alpha$ integral can be done if interchanged with the Fourier integrals of Eq.~(\ref{apm}). This gives a Dirac delta, which
 in turn allows one to evaluate the integral over $p_\p$. The result is,
\be
A(n;{\bf x}_i,{\bf x}_f;\p_i,\p_f)=\int_D \frac{\dd^4\theta}{(2\pi)^4} e^{i \theta_1  k^+_1+\theta_2 k^+_2+\theta_3
k^-_1+\theta_4  k^-_2} e^{i (\p_f-\p_i) \sqrt{2 \pi G n}\sqrt{(2n + 1)h(\theta_1,\theta_2)+ (2n - 1)h(\theta_3,\theta_4)}}~,
\ee
where ${\bf x}_f-{\bf x}_i={\bf k}^+ a^+_n+{\bf k}^- a^-_n$, \,  $h(\theta_1,\theta_2)=\cos \theta_1 + \cos \theta_2 + \cos 
(\theta_2-\theta_1)$
and the domain of integration is 
\be
D= \{(\theta_1,\theta_2,\theta_3,\theta_4) \in [0,2\pi)^4 \; / \, (2n + 1)h(\theta_1,\theta_2)+ (2n - 1)h(\theta_3,\theta_4) > 0 \}.
\ee

So far we have discussed histories amplitudes of the form $A(n_M,\ldots,n_0;{\bf x}_i,{\bf x}_f;\p_i,\p_f)$, where intermediate
 anisotropies are already summed over. Let us now discuss briefly how one would compute the amplitude for an individual
`colouring' of such a history, $A({\bf y}_M, n_M, {\bf x}_M, {\bf y}_{M-1}, n_{M-1}, \ldots, {\bf y}_0, n_0, {\bf x}_0)$.
The building blocks  in this case are the amplitudes $A_{n {\bf y} {\bf x}}(\tau)$ and $V_{n' {\bf x} n {\bf y}}$,
Eq.~(\ref{A-2}) and Eq.~(\ref{V-2}). The first  amplitude coincides with the $M=0$ case discussed above. We thus have that
 $A_{n {\bf y} {\bf x}}(\tau)$ vanishes unless ${\bf y}- 
{\bf x}={\bf k}^+ a^+_n+{\bf k}^- a^-_n$ in which case it is given by Eqs.~(\ref{mea}) and (\ref{apm}). On the other hand, the value
for $V_{n' {\bf x} n {\bf y}}$  can be easily read off from its definition:
it vanishes unless $n'=n\pm 1$ and ${\bf y}- {\bf x}$ lies in one of the following six points, $\{(a^{\pm}_n,0),(a^{\pm}_{n\pm 1/2},0),
(a^{\pm}_n,a^{\pm}_{n\pm 1/2}),(0,a^{\pm}_n),(0,a^{\pm}_{n\pm 1/2}),(a^{\pm}_{n\pm 1/2},a^{\pm}_n) \}$, in which case it takes the value
$- \pi G \sqrt{n(n \pm 1)}(2n \pm 1)$. After multiplying by the remaining factors in Eq.~(\ref{ampanis}), one has to perform the
same integrals as previously, in this case given in Eq.~(\ref{amphg2}) and Eq.~(\ref{amppath}).

\subsection{Vacuum case}

The presence of matter entered only at the very end of the construction. If we did not have matter at all, we could still follow the
same  procedure and arrive at the vacuum equivalent of Eq.~(\ref{physip-1}),
\be 
 ([n_f,{\bf x}_f] , [n_i, {\bf x}_i] ) = \sum_{M=0}^\infty \Big[ \sum_{n_{M-1},\ldots,n_{1}} A^{\text{vacuum}}(n_M,\ldots,
n_0;{\bf x}_i,{\bf x}_f) \Big]~.
\ee  
The difference between the matter and vacuum cases lies only in the form of the amplitudes, which in the vacuum case are formally
given by,
\be \label{ampvac}
A^{\text{vacuum}}(n_0,\ldots,n_M;{\bf x}_i,{\bf x}_f)= \sint_{-\infty}^{\infty}
 \dd{\alpha} \,  \bra {\bf x}_f | A(n_0,\ldots,n_M;\alpha) |{\bf x}_i \ket~.
\ee
These amplitudes can then be evaluated following the strategy given in the previous section, the only difference being the
absence of the final integral over $p_\p$. It is not obvious whether the integral in Eq.~(\ref{ampvac}) converges for all
 paths thus giving a meaningful expansion. Nevertheless, by looking at the generic behaviour of these integral, one finds
some evidence that it may converge. For instance, in the constant volume ($M=0$) path, Eq.~(\ref{ampvac}) gives
\be
A^{\text{vacuum}}(n;{\bf x}_i,{\bf x}_f)=\frac{1}{2 \pi G n}\int \frac{\dd^4\theta}{(2\pi)^4} e^{i \theta_1  k^+_1+
\theta_2  k^+_2+ \theta_3  k^-_1+\theta_4  k^-_2} \delta\left( (2n + 1)h(\theta_1,\theta_2)+ (2n - 1)
h(\theta_3,\theta_4) \right)~
\ee
which is clearly finite (at least for $n>0$). This is to be contrasted with the vacuum FRW case~\cite{ach2}, and the
example in~\cite{rv2}, where a regulator is required in order to render finite the otherwise divergent amplitudes,
even in this $M=0$ case. Further study of the vacuum case is in progress~\cite{Ed_Adam}.

\section{Projection to FRW}\label{s4}
At the end of section~\ref{s2}, we commented on the projection from  Bianchi~I to FRW. We show here that
when such projection is done at the level of the vertex expansion, Eq~(\ref{physip-1}), one recovers the FRW vertex
expansion of \cite{ach2}.

The structure in both cases is almost identical. One has a  sum over volume sequences $(n_M,\ldots,n_0)$, and each amplitude
is constructed by first obtaining a gravitational amplitude, and then performing the group averaging and scalar field
integrations. Thus, all that remains is to show that the amplitude $\bra {\bf x}_f |  A(n_M,\ldots,n_0;\t_M,\ldots,\t_1;\alpha)
 | {\bf x}_i \ket$ given in Eq.~(\ref{amphgt}) projects to the corresponding FRW one,
\begin{align} \label{ampfrw}
A^\frw(n_M,\ldots,n_0;\t_M,\ldots,\t_1;\alpha)= & \; e^{-i(\alpha - \tau_M) \Theta^\frw_{n_M n_M}}\,\,
 \Theta^\frw_{n_M n_{M-1}}\,\,
\times \nonumber\\
&\ldots\,\, e^{-i(\tau_2-\tau_1) \Theta^\frw_{n_1 n_1}}\,\,
 \Theta^\frw_{n_1 n_{0}}\,\, e^{-i\tau_1 
\Theta^\frw_{n_0 n_0}}~,
\end{align}
when summing over all possible values of ${\bf x}_f$. 

To show this, it is convenient to explicitly write the intermediate anisotropies in the amplitude $\bra {\bf x}_f |  A(n_M,\ldots,n_0;
\t_M,\ldots,\t_1;\alpha) | {\bf x}_i \ket$. As before, this is done by introducing complete basis $\mathbf{1}_\la= \sum_{{\bf x}_m}
| {\bf x}_m \ket \bra {\bf x}_m |$  and  $\mathbf{1}_\la= \sum_{{\bf y}_m}| {\bf y}_m \ket \bra {\bf y}_m |$ to the right and left
of the $A_{n_m}(\t_{m+1}-\t_m)$ operator in Eq.~(\ref{amphgt}). Calling ${\bf x}_f \equiv {\bf y}_M$ and ${\bf x}_i \equiv {\bf x}_0$
we have,
\begin{align}  \label{amphgtl}
   &\sum_{{\bf y}_M} \bra {\bf y}_M |  A(n_M,\ldots,n_0;\t_M,\ldots,\t_1;\alpha) | {\bf x}_0 \ket  = 
 \sum_{\substack{{\bf x}_1,\ldots,{\bf x}_M\\{\bf y}_0, \ldots,{\bf y}_{M}}}
  \bra {\bf y}_M | A_{n_M}(\alpha-\t_M) |{\bf x}_M \ket  \nonumber \\ 
   & \bra {\bf x}_M | V_{n_M n_{M-1}} |{\bf y}_{M-1}  \ket \bra {\bf y}_{M-1}|A_{n_{M-1}}(\t_M-\t_{M-1})|{\bf x}_{M-1}
  \ket \ldots \bra {\bf x}_1 |  V_{n_1 n_0}|{\bf y}_{0} \ket \bra {\bf y}_{0}|A_{n_0}(\t_1)|{\bf x}_0 \ket~.
\end{align}
We now use the translation invariance of the operators to write each matrix elements in Eq.~(\ref{amphgtl}) as $\bra {\bf x}'| f |
{\bf x} \ket = \bra {\bf x}'-{\bf x}| f | 0 \ket$ where $f$ is either the $A$ or $V$ operators. We then change the
summation variables to ${\bf y}'_m={\bf y}_m-{\bf x}_m$ and ${\bf x}'_m={\bf x}_m-{\bf y}_{m-1}$.
The different summations in Eq.~(\ref{amphgtl}) then decouple giving,
\begin{align} \label{amphgtl2}
   &\sum_{{\bf y}_M} \bra {\bf y}_M |  A(n_M,\ldots,n_0;\t_M,\ldots,\t_1;\alpha) | {\bf x}_0 \ket  =  \left(\sum_{{\bf y}'_M}
  \bra {\bf y}'_M | A_{n_M}(\alpha-\t_M) |0 \ket\right) \left(\sum_{{\bf x}'_M} \bra {\bf x}'_M | V_{n_M n_{M-1}} |0  \ket \right) 
  \nonumber \\   &  \left(\sum_{{\bf y}'_{M-1}} \bra {\bf y}'_{M-1}|A_{n_{M-1}}(\t_M-\t_{M-1})|{\bf x}_{M-1} \ket \right) \ldots
  \left(\sum_{{\bf x}'_1} \bra {\bf x}'_1 | V_{n_1 n_0}|0 \ket \right) \left(\sum_{{\bf y}'_0} \bra {\bf y}'_{0}|A_{n_0}(\t_1)|0
 \ket\right)~.
\end{align}
Comparing Eq.~(\ref{amphgtl2}) and Eq.~(\ref{ampfrw}) we see that our task reduces to showing that
 $\sum_{\bf x} \bra {\bf x} | A_{n}(\t) |0 \ket = e^{-i\t \Theta^{FRW}_{n n}}$ and $\sum_{\bf x}
\bra {\bf x} | V_{n' n} |0 \ket = \Theta^{FRW}_{n' n}$. That this
 is so can be seen as a direct consequence of the result in~\cite{awe}. Let us nevertheless show it explicitly.

For the $V$ term, it suffices to look at $V_{n\pm1 n}$. The operator, given
in Eq.~(\ref{offdiag}) consist in an overall constant times six different translations in the ${\bf x}$ plane. Each term thus
will pick up a single ${\bf x}$ from the sum. For instance, the first term gives a nonzero value only for ${\bf x}=(a^\pm_n,0)$,
in which case it gives a contribution of $- \pi G \sqrt{n(n \pm 1)}(2n \pm 1)$. We then conclude that
\be \label{vfrw}
\sum_{{\bf x}} \bra {\bf x}| V_{n\pm1 n} |0 \ket = - 6 \pi G \sqrt{n(n \pm 1)}(2n \pm 1) = \Theta^{FRW}_{n\pm 1 n}~,
\ee
as required.

For the $A_n(\t)$ term we have
\begin{align} \label{afrw}
& \sum_{\bf x} \bra {\bf x} | A_{n}(\t) |0 \ket = \nonumber \\
 & \sum_{{\bf k}^+,{\bf k}^- \in \Z^2} \bra {\bf k}^+|A^+_n(\t)|0 \ket \bra {\bf k}^-|A^-_n(\t)|0 \ket = \nonumber \\
& e^{-i \t 24 \pi G n^2} = e^{-i \t \Theta^{FRW}_{n n}}~,
\end{align}
where in going from the first to second line, we used Eq.~(\ref{mea}). In going from the second
to third line, we used Eq.~(\ref{apm}) and the identity $\sum_{k \in \Z} e^{i k \theta}/2\pi = \delta(\theta)$
to directly evaluate the Fourier integrals. Using Eqs.~(\ref{vfrw}) and (\ref{afrw}) we then have
\be
\sum_{{\bf x}_f} \bra {\bf x}_f |  A(n_M,\ldots,n_0;\t_M,\ldots,\t_1;\alpha) | {\bf x}_i \ket = A^\frw(n_M,\ldots,n_0;\t_M,\ldots,\t_1;\alpha)~,
\ee
which implies
\be
\sum_{{\bf x}_f} A(n_M,\ldots,n_0;{\bf x}_i,{\bf x}_f;\p_i,\p_f) =  A^\frw(n_M,\ldots,n_0;\p_i,\p_f)~.
\ee
Thus we see that the vertex expansion for our Bianchi model, Eq.~(\ref{physip-1}) projects down to the vertex expansion
of the FRW model, order by order.

\section{Discussion}\label{s5}

Recently it has been shown~\cite{ach1} that one can take the Loop Quantum version of FRW cosmology
and expand it as a sum over volume transition of amplitudes compatible with given initial and final states i.e.\
that the cosmological model of Loop Quantum Gravity can be re-written in terms of a sum over amplitudes, analogous to
the spin foam approach. This sum over transition amplitudes is produced as a perturbation expansion of the
constraint operator of LQC, thus linking perturbative dynamics of LQC to (the analogue of) spin foams.
This analogue provided a useful link between the two theories, however because there is only
one dynamic parameter in FRW cosmologies -- the volume -- the system has no analogue of the spin labels. In this
paper we have extended the approach of~\cite{ach1} to the Bianchi~I cosmological model, which, in addition to volume,
has anisotropic degrees of freedom. We have shown that it is again possible to expand the dynamics of the
model in terms of sums of amplitudes over volume transitions compatible with initial and final states. The additional
anisotropic degrees of freedom of this model are analogous to the spin labels of spin networks, thus significantly
improving the analogue to spin foams.

The analogue remains at a formal level however, because one cannot directly associate the amplitudes with the 
changing of an underlying spin network. Despite this the association of the anisotropic degrees of freedom with
the spin labels is well motivated by the fact that in LQC they give the area (of the fiducial cell), which is
precisely the role played by the spin labels (and edges) in a spin network. In addition to showing that the resulting
summation over `spin' labels is finite, we show that the projection to the FRW system occurs order by order
in the expansion, thus recovering the results of~\cite{ach1}.

Finally, although spin foams are typically taken to have spin changes only at vertices, it is generally
expected that spin dynamics in the absence of graph changing vertices will play an important role in the final
theory~\cite{freidel}. More precisely, that the action of the full constraint is non-trivial, even in the absence of vertices
and hence that the amplitude for each vertex-free segment of the spin foam will be non-diagonal in the spin labels. In
the analogue produced here we show that indeed the `spin' changing amplitude is non-trivial, even in the absence of volume
changing `vertices'. Thus our full expansion is the analogue of a generalization of spin foams, allowing for
`spin' dynamics.

\section*{Acknowledgments}

We would like to thank Abhay Ashtekar and Edward Wilson-Ewing for illuminating discussions.
This work was supported in part by NSF grants PHY0748336,
PHY0854743, The George A.\ and Margaret M.~Downsbrough Endowment and
the Eberly research funds of Penn State.

\end{document}